\documentclass[useAMS,usenatbib,fleqn]{mn2e}

\usepackage{graphicx}
\usepackage{float}
\usepackage{epstopdf}
\usepackage{amsmath}
\usepackage{amssymb}
\usepackage[labelformat=empty,labelsep=none]{subcaption}
\usepackage{epsfig}
\usepackage{times}
\usepackage{multirow}
\usepackage{color}

\def\gtsim {\gtrsim}   
\def\ltsim {\lesssim}   
\def\ldm {\Delta_{\rm logM}}

\newcommand{\dummytitle}[1]{}

\newcommand{\msun}{{\,\rm M}_\odot}

\defcitealias{pt14}{TK14}
\defcitealias{pt15a}{TK15a}
\defcitealias{pt15b}{TK15b}

\newcommand{\apm}[2]{^{+#1}_{-#2}}

\title[The Star Formation Rate in Simulated Galaxies]{Star Formation in Simulated Galaxies: Understanding the Transition to Quiescence at $3\times10^{10}\msun$}
\author[P.~Taylor, C.~Federrath, and C.~Kobayashi]{Philip~Taylor$^1$\thanks{E-mail: philip.1.taylor@anu.edu.au}, Christoph~Federrath$^1$, and Chiaki~Kobayashi$^2$\\
$^1$Research School of Astronomy and Astrophysics, Australian National University, Canberra, ACT 2611, Australia\\
$^2$Centre for Astrophysics Research, Science and Technology Research Institute, University of Hertfordshire, AL10 9AB, UK}

\begin{document}

\date{Accepted  Received ; in original form}

\pagerange{\pageref{firstpage}--\pageref{lastpage}} \pubyear{}

\maketitle

\label{firstpage}

\begin{abstract}
Star formation in galaxies relies on the availability of cold, dense gas, which, in turn, relies on factors internal and external to the galaxies.
 {In order to provide a simple model for how star formation is regulated by various physical processes in galaxies,} we analyse data at redshift $z=0$ from a hydrodynamical cosmological simulation that includes prescriptions for star formation and stellar evolution, active galactic nuclei (AGN), and their associated feedback processes.
This model can determine the star formation rate (SFR) as a function of galaxy stellar mass, gas mass, black hole mass, and environment.
We find that gas mass is the most important quantity controlling star formation in low-mass galaxies, and star-forming galaxies in dense environments have higher SFR than their counterparts in the field.
In high-mass galaxies, we find that black holes more massive than $\sim10^{7.5}\msun$ can be triggered to quench star formation in their host; this mass scale is emergent in our simulations.
Furthermore, this black hole mass corresponds to a galaxy bulge mass $\sim2\times10^{10}\msun$, consistent with the mass at which galaxies start to become dominated by early types ($\sim3\times10^{10}\msun$, as previously shown in observations by \citeauthor{kauffmann03}).
 {Finally, we demonstrate that our model can reproduce well the SFR measured from observations of galaxies in the GAMA and ALFALFA surveys.}

\end{abstract}



\section{Introduction}
\label{sec:intro}

Understanding how galaxies evolve is a central problem in astronomy.
It has been recognised for some time that energy feedback from stars and super-massive black holes (BH) in active galactic nuclei (AGN) is responsible for suppressing star formation in low- and high-mass galaxies, respectively.
The transition between these two regimes occurs at a galaxy stellar mass $M_*\sim 3\times10^{10}\msun$ \citep[e.g.,][]{kauffmann03,baldry06}.
The aim of this study is to improve our understanding of the physical processes that influence star formation in all galaxies and give rise to this transition scale.

Stars are formed according to an initial mass function (IMF), which describes the distribution of stellar masses formed \citep[e.g.,][]{salpeter55,chabrier03,kroupa08,kroupa13}.
Very massive, but short-lived, stars explode as core-collapse supernovae, heating the surrounding interstellar medium (ISM), and enriching it with metals.
Lower mass stars can also produce supernovae on longer (Gyr) timescales via white dwarf or neutron star progenitors.
The injection of energy into the ISM by both stellar feedback \citep[e.g.,][]{heckman00,pettini00,ohyama02} and AGN feedback \citep{lynds67,kraft09,feruglio10,cicone12,tombesi13,teng14} can suppress further star formation on galactic scales \citep[see also, e.g.,][who discuss the possibility of AGN-induced star formation on small scales]{bicknell00,silk13,zubovas13,shabala15,bieri16}.
Stellar and AGN feedback can drive galactic outflows, which have been reproduced in cosmological simulations \citep[e.g.,][]{pt15b}.
By quenching star formation, AGN feedback in massive galaxies causes them to have lower specific star formation rates, enhanced [$\alpha$/Fe], and redder colours than lower mass galaxies, which may lead to the observed downsizing phenomenon \citep{cowie96,juneau05,bundy06,stringer09}.

Stars form in the densest regions of giant molecular clouds in the ISM \citep{elmegreen04,maclow04,mckee07,cf12,padoan14}.
AGN activity, too, relies on the accretion of gas onto the central super-massive BH.
The availability of gas for both star formation and AGN therefore plays an important role in the subsequent evolution of a galaxy.

 {Most previous theoretical work has sought to explain observed star formation rates (SFRs) as a fraction of the molecular gas mass per (density-dependent) free-fall time \citep{krumholz12,cf13,salim15,semenov16}.
Such studies can explain local SFRs on the scale of giant molecular clouds, but a full understanding of the processes that affect the global galactic SFR is lacking.
Observational data showed that there are two important, empirical relations: the Kennicutt-Schmidt law \citep{schmidt59,kennicutt89,kennicutt12} and the star formation main sequence \citep[SFMS; e.g.,][]{elbaz11,zahid12,renzini15}.
Most previous theoretical work has sought to explain the former, but a full understanding of the latter is lacking.
\citet{feldman16} use high-resolution simulations from the Feedback in Realistic Environments (FIRE) project \citep{hopkins14} to study the processes governing the SFR of galaxies at $z\sim2$, but do not explore the relative importance of these processes.}

Galaxies form from massive gas clouds at high redshift, but can both gain and lose gas over the course of their life.
Simulations based on the CDM cosmology predict large-scale flows of cold gas along filaments in the cosmic web \citep[e.g.,][]{dekel09,cen14b} that can efficiently supply a galaxy with gas without being shock-heated to the virial temperature \citep[e.g.,][]{birnboim16}.
Groups and clusters are found at more massive nodes in the web, and can be fed by a greater number of filaments.

Gas is also transported into galaxies via mergers.
So-called `wet' (two gas-rich galaxies) and `damp' (one gas-rich and one gas-poor galaxy) mergers can provide a galaxy with gas for both star formation and AGN activity, as well as potentially altering the morphology.
During a merger, gas can be efficiently transported to the centre of the new galaxy, providing fuel for star formation and AGN activity \citep{toomre72,combes90,mihos94b,mihos96,barnes96,hopkins08}.
Major mergers, in which the galaxies have similar masses, are typically thought to be the main drivers of galaxy evolution (in terms of morphology, SFR, and AGN activity) in the local universe \citep[e.g.,][]{darg10b}, though minor mergers contribute significantly to the cosmic star formation budget at low redshift \citep[e.g.,][]{kaviraj14a}.
In cosmological simulations, both major and minor mergers occur according to the hierarchical clustering of galaxy halos.

Note that, as well as the feedback processes described above, galaxies can lose gas when falling into a cluster.
Tidal interactions with cluster members and the dark matter halo itself can have an effect \citep[e.g.,][]{moore96}, and ram pressure stripping of the interstellar medium by the hot intra-cluster medium can remove much of the halo gas \citep[e.g.,][]{gunn72,dressler83,gavazzi95,boselli06}.
These effects are not well reproduced in smoothed particle hydrodynamics (SPH) simulations.

It is clear that there are a number of complex and competing processes that alter the amount of gas in a galaxy that can form stars or fuel BHs.
What is less apparent is which of these processes is more important, and in what circumstances, in determining the SFR?
In this paper, we aim to quantitatively answer this using cosmological hydrodynamical simulations that self-consistently solve the relevant physics.
Section \ref{sec:sims} gives a brief overview of the simulations analysed.
In Section \ref{sec:model}, we describe the analysis methodology including model formulae, and determine the model parameters in Section \ref{sec:model_disc}.
 {Section \ref{sec:discussion} presents a discussion of the results, and in Section \ref{sec:obs} we test our model against observational data.}
Finally, we give our conclusions in Section \ref{sec:conc}.


\section{Simulations}
\label{sec:sims}

The simulations used in this paper were introduced in \citet{pt15a}; they are a pair of cosmological, chemodynamical simulations, one of which includes a model for AGN feedback \citep{pt14}, but otherwise have indentical initial conditions and physics.
Our simulation code is based on the SPH code {\sc gadget-3} \citep{springel05gadget}, updated to include:  star formation \citep{ck07}, energy feedback and chemical enrichment from supernovae \citep[SNe II, Ibc, and Ia,][]{ck04,ck09} and hypernovae \citep{ck06,ck11a}, and asymptotic giant branch (AGB) stars \citep{ck11b}; heating from a uniform, evolving UV background \citep{haardt96}; metallicity-dependent radiative gas cooling \citep{sutherland93}; and a model for BH formation, growth, and feedback \citep{pt14}, described in more detail below.
We use the IMF of stars from \citet{kroupa08} in the range $0.01-120\msun$, with an upper mass limit for core-collapse supernovae of $50\msun$.

The initial conditions for both simulations consist of $240^3$ particles of each of gas and dark matter in a periodic, cubic box $25\,h^{-1}$ Mpc on a side, giving spatial and mass resolutions of $1.125\,h^{-1}$ kpc and $M_{\rm DM}=7.3\times10^7\,h^{-1}\msun$, $M_{\rm g}=1.4\times10^7\,h^{-1}\msun$, respectively. 
We employ a WMAP-9 $\Lambda$CDM cosmology \citep{wmap9} with $h=0.7$, $\Omega_{\rm m}=0.28$, $\Omega_\Lambda=0.72$, $\Omega_{\rm b}=0.046$, and $\sigma_8=0.82$.

BHs form from gas particles that are metal-free and denser than a specified critical density, mimicking the most likely formation channels in the early Universe via direct collapse of a massive gas cloud \citep[e.g.,][]{bromm03,koushiappas04,agarwal12,becerra15,regan16a,hosokawa16} or as the remnant of Population {\sc III} stars \citep[e.g.,][]{madau01,bromm02,schneider02}.
The BHs grow through gas accretion and mergers.
The accretion rate is estimated assuming Bondi-Hoyle accretion:
\begin{equation}\label{eq:macc}
	\dot{M}_{\rm acc} = \frac{4\pi G^2 M^2_{\rm BH}\rho}{\left(c^2_s + v^2\right)^{3/2}},
\end{equation}
where $M_{\rm BH}$ is the mass of the BH, $\rho$ the local gas density, $c_s$ the local sound speed, and $v$ the speed of the BH relative to the local gas particles.
We limit the accretion rate to the Eddington rate, given by 
\begin{equation}\label{eq:medd}
	\dot{M}_{\rm Edd} = \frac{4\pi G M_{\rm BH} m_{\rm p}}{\epsilon_{\rm r}\sigma_{\rm T}c} \equiv \frac{M_{\rm BH}}{t_{\rm Sal}},
\end{equation}
where $m_{\rm p}$ is the proton mass, $\sigma_{\rm T}$ is the Thompson cross section, $\epsilon_{\rm r}$ denotes the radiative efficiency of the BH, which we set to $0.1$, and $t_{\rm Sal}\approx 45\,{\rm Myr}$ is the Salpeter time.
Two BHs merge if their separation is less than the gravitational softening used, and their relative speed is less than the local sound speed.
A fraction of the energy liberated by gas accretion is coupled to neighbouring gas particles in a purely thermal form.
 


\section{A new simple model for the SFR}
\label{sec:model}

\begin{figure}
	\centering
	\includegraphics[width=0.48\textwidth,keepaspectratio]{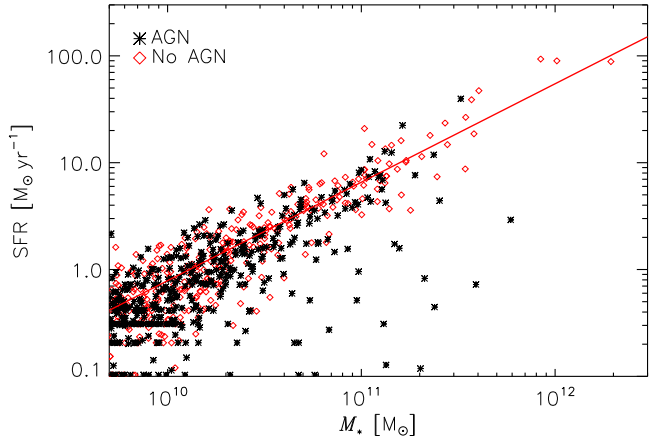}
	\caption{Star formation main sequence from our simulations.
	Red diamonds are for the simulation with AGN physics turned off, and black stars for the simulation including all physics.
	With AGN feedback included, some galaxies move off the main sequence to lower SFR.
	The solid red line shows the fit to the data from the simulation without AGN feedback (equation \eqref{eq:sfms}).}
	\label{fig:sfms}
\end{figure}
The main aim of this paper is to gain a quantitive understanding of the relative importance of how various physical processes affect the SFR of galaxies at the present day.
We start from the SFMS, which relates the SFR and stellar mass of galaxies.
Observations show that star-forming, late-type galaxies (LTG) form a narrow sequence with more massive galaxies having higher SFR \cite[e.g.,][]{elbaz11,zahid12,renzini15}, while early-type galaxies (ETG) occupy a region of parameter space below this sequence \citep[e.g.,][]{wuyts11,renzini15}.
In our simulations, AGN feedback is responsible for quenching star formation and moving galaxies off the SFMS \citep{pt16}.
Fig. \ref{fig:sfms} shows our simulated SFMS both with (black stars) and without (red diamonds) AGN feedback.
The solid red line in Fig. \ref{fig:sfms} shows the linear bisector fit to the data from the simulation without AGN feedback, given by
\begin{equation}\label{eq:sfms}
	\log{\rm SFR}_{\rm fit} = -9.32 + 0.92\log M_*,
\end{equation}
where $M_*$ is the galaxy stellar mass (see \citet{pt16} for a detailed comparison between our simulated SFMS and observational data).
Using only the data from this simulation to obtain equation \eqref{eq:sfms} ensures that all galaxies are star-forming and lie on the SFMS, giving a relationship with no AGN contamination.

However, as discussed in Section \ref{sec:intro}, other factors, such as the mass of gas in a galaxy, and the environment the galaxy exists in, may also affect the SFR.
Feedback from star formation, supernovae, and AGN activity can remove gas from galaxies \citep{pt15b}, depending on the galaxy mass and strength of feedback.
Although the feedback energy in our AGN model is calculated from the instantaneous accretion rate, the current mass of gas in the galaxy likely depends more on the BH mass, since this reflects the entire accretion history of the BH.
Similarly, the total stellar mass better reflects the integrated influence of stellar feedback on the gas than the current SFR.

Therefore, in order to quantify the relative importance of these processes, we suggest the following simple model to describe SFR in all galaxies:
\begin{equation}\label{eq:10tomodel}
	{\rm SFR} = A \times M_*^\alpha \times s_5^\beta \times M_{\rm gas}^\gamma \times f(M_{\rm BH}),
\end{equation}
where $s_5$ is the 3-dimensional 5$^{\rm th}$-nearest neighbour distance, which we showed in Taylor \& Kobayashi (in prep.) to be a good measure of environmental density \citep[though numerous measures of environmental density have been proposed; see e.g.,][]{haas12}, $M_{\rm gas}$ is the mass of gas within the galaxy, and $A$ is a constant.
The function $f(M_{\rm BH})$ will be discussed in detail below.

To perform the fitting of our simulated data to this model, it is convenient to work with the logarithm\footnote{Throughout this paper we adopt the standard notation $\log x \equiv \log_{10}x$ and $\ln x \equiv \log_ex$.} of equation \eqref{eq:10tomodel}:
\begin{equation}\label{eq:model}
\begin{split}
	\log {\rm SFR} =  C & + \alpha\log\left(\frac{M_*}{10^{10}\msun}\right) + \beta\log\left(\frac{s_5}{10^2{\rm kpc}}\right)  \\
	 & + \gamma\log\left(\frac{M_{\rm gas}}{10^7\msun}\right) + \log f \left(\frac{M_{\rm BH}}{10^8\msun}\right).
\end{split}
\end{equation}
The coefficients $C$, $\alpha$, $\beta$, and $\gamma$ are constants that will be determined by fitting (note that $\alpha$, $\beta$ and $\gamma$ are the scaling exponents for $M_*$, $s_5$, and $M_{\rm gas}$, respectively, as defined in equation \eqref{eq:10tomodel}).
$M_*$, $s_5$, $M_{\rm gas}$, and $M_{\rm BH}$ are normalised by representative quantities, such that the distribution of $\log \left(M_*/10^{10}\msun\right)$ (and similarly for $s_5$, $M_{\rm gas}$, and $M_{\rm BH}$) has a modal value of about zero; this too is a fitting convenience, and affects only the value of fitted parameter $C$.

\begin{figure}
	\centering
	\includegraphics[width=0.48\textwidth,keepaspectratio]{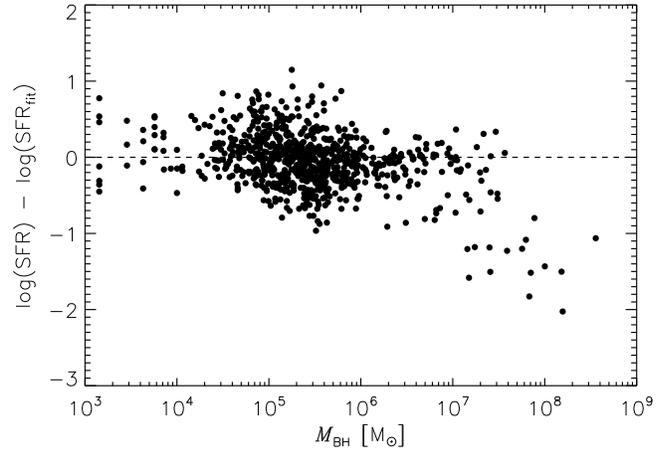}
	\caption{Difference between measured SFR from the galaxies in the simulation with AGN feedback, and the SFR expected given the stellar masses of those galaxies and the fit to the SFMS from the simulation without AGN feedback.}
	\label{fig:dsfr_mb}
\end{figure}

\begin{figure}
	\centering
	\includegraphics[width=0.48\textwidth,keepaspectratio]{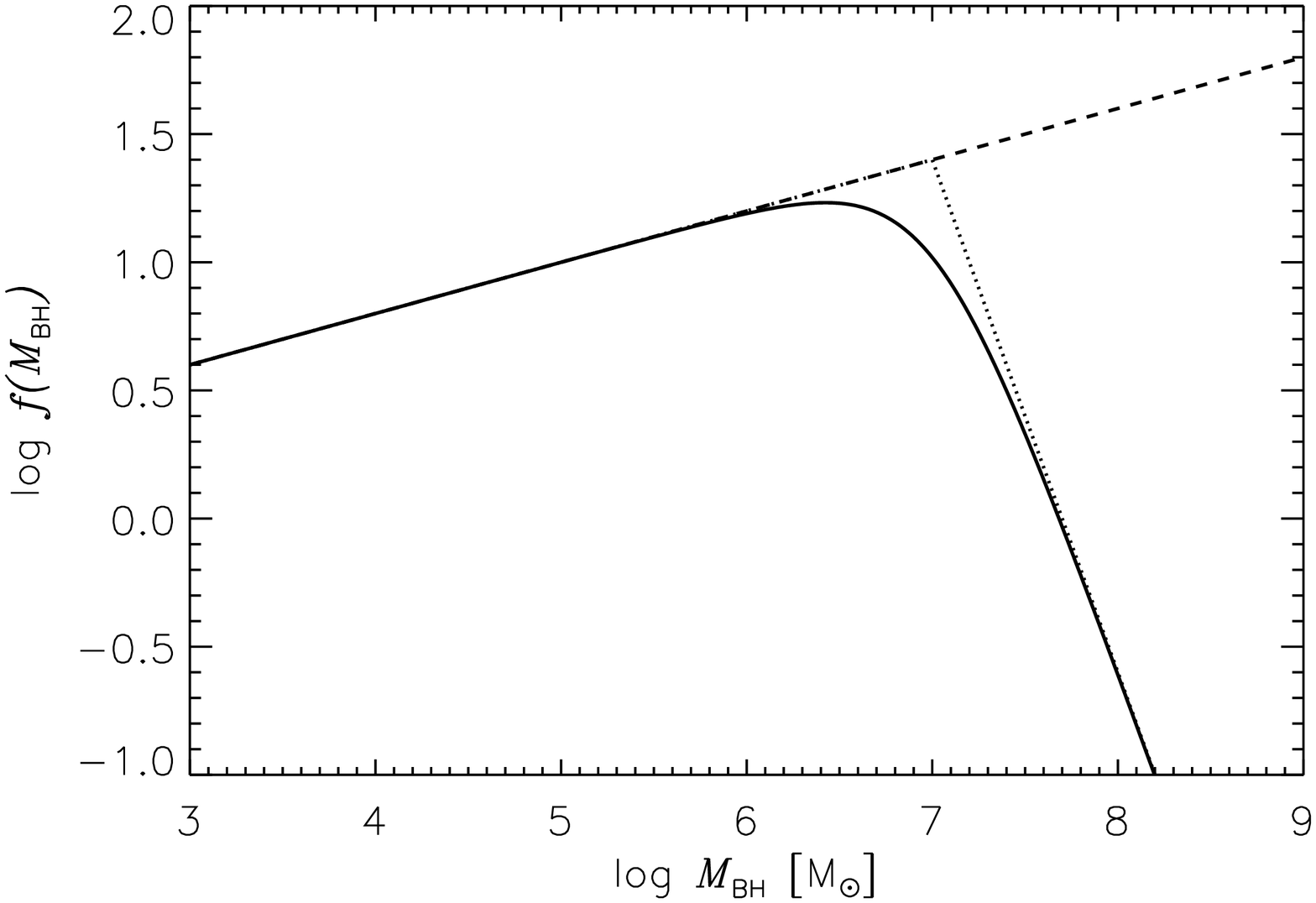}
	\caption{Examples of the three functions given by Models 1-3, with arbitrary example parameters $k_1=0.2$, $k_2=-2$, $\log M_{\rm b}=7$, and $\ldm=1$.
	Note that these values are purely illustrative; measured values are presented in Section \ref{sec:model_disc}.}
	\label{fig:curves}
\end{figure}

The function $f(M_{\rm BH})$ in equations \eqref{eq:10tomodel} and \eqref{eq:model} reflects the fact that BHs must grow sufficiently massive before their feedback energy can not be efficiently radiated away.
This is illustrated in Fig. \ref{fig:dsfr_mb}, which shows the residuals of SFR in the simulation with AGN compared with the SFR estimated from equation \eqref{eq:sfms}, as a function of BH mass.
There are two clear regimes: at $M_{\rm BH}\ltsim10^7\msun$, the residuals are distributed around $0$, while at $M_{\rm BH}\gtsim10^7\msun$ the residuals decrease with increasing $M_{\rm BH}$, and the SFR can be as much as two orders of magnitude below the SFMS.
This suggests the following form for $f(M_{\rm BH})$:
\begin{equation}\label{eq:model1}
	\log f_1(M_{\rm BH}) = 
	\begin{cases}
		k_1\log M_{\rm BH} & M_{\rm BH}< M_{\rm b}, \\
		k_2\log M_{\rm BH} + (k_1-k_2)\log M_{\rm b} & M_{\rm BH}\ge M_{\rm b},
	\end{cases}
\end{equation}
where $k_1$ and $k_2$ are constant, and $M_{\rm b}$ is a break mass at which $f(M_{\rm BH})$ turns over\footnote{Given the normalisation of $M_{\rm BH}$ to $10^8\msun$ in equation \eqref{eq:model}, $M_{\rm b}$ is also measured relative to $10^8\msun$. In subsequent sections, the absolute value of $M_{\rm b}$ will be given, with the factor $10^8$ taken into account.}.
In Section \ref{sec:model_disc}, we also test the following forms of $f(M_{\rm BH})$:
\begin{equation}\label{eq:model2}
	\log f_2(M_{\rm BH}) = k_1\log M_{\rm BH},
\end{equation}
\begin{equation}\label{eq:model4}
	\begin{split}
		\log f_3 & (M_{\rm BH})  = \frac{1}{2}(k_1+k_2)\log M_{\rm BH} \\ 
		&+ \frac{1}{2}(k_2-k_1) \ldm \ln\left(\cosh\left(\frac{\log M_{\rm BH}-\log M_{\rm b}}{\ldm}\right)\right).
	\end{split}
\end{equation}
Function $f_2$ is the special case of $f_1$ in which $k_2=k_1$, i.e. there is no break.
$f_3$ is a continuous, smooth analogue to $f_1$, in which the gradient varies smoothly from $k_1$ to $k_2$, with the additional parameter $\ldm$ controlling the width over which the transition occurs.
Equation \eqref{eq:model4} reproduces $f_1$ in the limit $\ldm\rightarrow 0$.
The constant of integration\footnote{$f_3$ reproduces $f_1$ to within a constant. Matching at $\log M_{\rm BH}\rightarrow\pm\infty$ gives this constant as $\frac{1}{2}(k_2-k_1)(\ldm \ln2 - \log M_{\rm b})$. In practice, we absorb this into $C$ to avoid unnecessary correlation between $C$ and the other parameters.} that would appear in equation \eqref{eq:model4} can simply be absorbed by $C$ in equation \eqref{eq:model}.
We show in Fig. \ref{fig:curves} an example of each of the three functions described above.

Values of $C$, $\alpha$, $\beta$, $\gamma$, $k_1$, $k_2$, $\log M_{\rm b}$, and $\ldm$ are found using the \texttt{amoeba} routine \citep{amoeba}, minimising the quantity $\sum_i\left(\log {\rm SFR}_i - \log {\rm SFR}_{i,{\rm fit}}\right)^2$.
In order to estimate uncertainties on the fitted parameters we employ a bootstrap resampling technique whereby the parameters are re-derived for a random selection of the data, with repeats, the same size as the original dataset.
We do this for $5\times10^6$ resamplings, in order to fully sample the parameter space of the models with the most free parameters.

\section{Model Fits}\label{sec:model_disc}

\begin{table}
\caption{Fitted parameters of the models presented in Section \ref{sec:model}.
The parameters denote: $C$ -- normalisation; $\alpha$ -- coefficient of $\log M_*$; $\beta$ -- coefficient of $\log s_5$, $\gamma$ -- coefficient of $\log M_{\rm gas}$; $k_1$ -- coefficient of $\log M_{\rm BH}$ (low mass); $k_2$ -- coefficient of $\log M_{\rm BH}$ (high mass); $\log M_{\rm b}$ -- break mass separating low- and high-mass BHs; $\ldm$ -- width of transition from low- to high-mass BHs.
Three values for each parameter are given for each model: the first is the best-fit value from the simulated data; the second is the mean and standard deviation from the full bootstrap parameter distributions; and the third is the modal value of the distributions with asymmetric errors denoting the $16^{\rm th}$ and $84^{\rm th}$ percentiles.
The distribution of $\ldm$ in model 3 is bimodal, so asymmetric error bars are unavailable; this value is marked $^\dagger$.}
	\begin{tabular}[width=0.48\textwidth]{crrr}
		$x$ & \multicolumn{1}{c}{Best Fit} & \multicolumn{1}{c}{$\bar{x} \pm \sigma_x$} & \multicolumn{1}{c}{Mode $\apm{}{}$ distribution width} \\
		\hline
		\multicolumn{4}{c}{Model 1 (equation \eqref{eq:model1})} \\
		\hline
		$C$ & $-2.50$ & $-2.59\pm0.18$ & $-2.58\apm{0.18}{0.18}$ \\
		$\alpha$ &  $0.21$ & $0.21\pm0.03$ & $0.21\apm{0.03}{0.03}$ \\
		$\beta$ & $-0.17$ & $-0.15\pm 0.04$ & $-0.16\apm{0.04}{0.03}$ \\
		$\gamma$ & $1.03$ & $1.04\pm0.05$ & $1.04\apm{0.05}{0.05}$ \\
		$k_1$ & $0.00$ & $-0.02\pm0.03$ & $-0.01\apm{0.03}{0.04}$ \\
		$k_2$ & $-1.11$ & $-1.99\pm3.13$ & $-1.12\apm{0.52}{0.82}$ \\
		$\log M_{\rm b}$ & $7.31$ & $7.79\pm0.64$ & $7.29\apm{0.46}{0.15}$ \\
		\hline
		\multicolumn{4}{c}{Model 2 (equation \eqref{eq:model2})} \\
		\hline
		$C$ & $-2.72$ & $-2.72\pm0.15$ & $-2.72\apm{0.15}{0.14}$ \\
		$\alpha$ & $0.21$ & $0.21\pm0.03$ & $0.21\apm{0.03}{0.03}$ \\
		$\beta$ & $-0.12$ & $-0.12\pm0.04$ & $-0.12\apm{0.04}{0.03}$ \\
		$\gamma$ & $1.05$ & $1.05\pm0.05$ & $1.04\apm{0.06}{0.05}$ \\
		$k_1$ & $-0.05$ & $-0.05\pm0.02$ & $-0.04\apm{0.02}{0.03}$ \\
		\hline
		\multicolumn{4}{c}{Model 3 (equation \eqref{eq:model4})}\\
		\hline
		$C$ & $-3.12$ & $-3.08\pm0.19$ & $-3.08\apm{0.19}{0.18}$ \\
		$\alpha$ & $0.21$ & $0.22\pm0.03$ & $0.21\apm{0.03}{0.03}$ \\
		$\beta$ & $-0.16$ & $-0.16\pm0.03$ & $-0.16\apm{0.03}{0.03}$ \\
		$\gamma$ & $1.05$ & $1.05\pm0.05$ & $1.06\apm{0.05}{0.05}$ \\
		$k_1$ & $0.04$ & $0.06\pm0.05$ & $0.08\apm{0.03}{0.08}$ \\
		$k_2$ & $-1.19$ & $-1.25\pm0.85$ & $-0.82\apm{0.31}{0.53}$ \\
		$\log M_{\rm b}$ & $7.68$ & $7.66\pm0.15$ & $7.58\apm{0.14}{0.07}$ \\
		$\ldm$ & $1.22$ & $1.55\pm0.76$ & $2.06^\dagger\,\,\,\,\,\,\,\,\,\,\,$ \\
	\end{tabular}
\label{tab:sfrfits}
\end{table}
\begin{figure*}
	\centering
	\includegraphics[width=\textwidth,keepaspectratio]{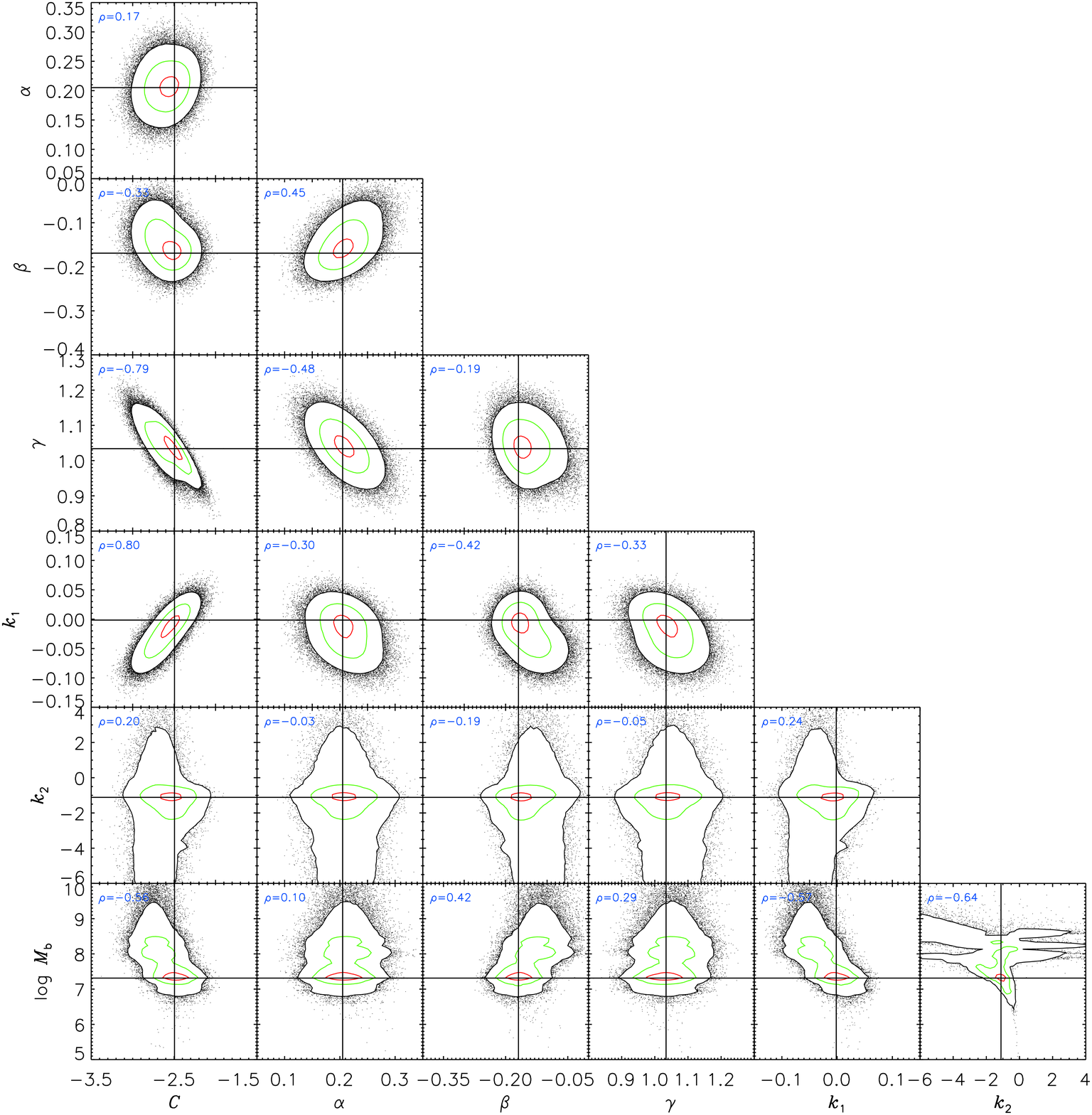}
	\caption{Parameter distributions for $f_1$ (equation \eqref{eq:model1}).
	Contours enclose 95, 68, and 16 per cent of points.
	The correlation coefficient, $\rho$, for each pair of parameters is shown in the upper left of each panel.}
	\label{fig:model1covarsfr}
\end{figure*}
Fitting the SFR of our simulated galaxies using the method described in Section \ref{sec:model} to the models described in equations \eqref{eq:model}-\eqref{eq:model4} results in the parameters given in Table \ref{tab:sfrfits}; the full parameter distributions for $f_1$ are shown in Fig. \ref{fig:model1covarsfr}.
The uncertainties are derived from the bootstrap resampling procedure described in Section \ref{sec:model}; the first column of Table \ref{tab:sfrfits} gives the best-fit parameters to the simulated data, the second gives the mean and standard deviation from the bootstrap distributions, and third shows the modal value with asymmetric errors denoting the $16^{\rm th}$ and $84^{\rm th}$ percentiles.
For all three models, the values of $C$, $\alpha$, $\beta$, and $\gamma$ are consistent with one another, and are well-constrained by the data.

We find that SFR depends most strongly on the amount of gas in a galaxy, scaling linearly with $M_{\rm gas}$ ($\gamma\approx1$).
Additionally, galaxies with large stellar mass and high environmental density (low $s_5$) show greater SFR, with the dependence on galaxy mass being the more important (i.e., $\alpha > |\beta|$).
However, with $f_1$ and $f_3$, this is tempered by the strong negative correlation between SFR and $M_{\rm BH}$ in galaxies with BHs more massive than $M_{\rm b}$ (i.e., $k_2$ is large and negative).
Such galaxies tend to be found in the densest environments, at the centre of clusters, implying that satellite galaxies within the cluster have the highest SFRs, while field galaxies tend to have lower SFRs at a given stellar mass.
This is consistent with the observational result of \citet{koyama13}, who found that the SFR of star-forming galaxies increases with environmental density \citep[see also][who find weak or no dependence of SFR on environment for star-forming galaxies]{peng10,wijesinghe12}.
In all models, star formation in galaxies with low-mass BHs is not strongly affected, as indicated by the value $k_1\sim0$, i.e. SFR follows the SFMS for $M_{\rm BH}\ltsim M_{\rm b}$ (see Fig. \ref{fig:dsfr_mb}).

The relatively large uncertainties on the values of both $k_2$ $\log M_{\rm b}$ in $f_1$ are due to their distributions, shown in Fig. \ref{fig:model1covarsfr}, being highly extended, and, in the case of $\log M_{\rm b}$, nearly bimodal.
In addition to the main peak around $\log M_{\rm b}\approx7.3$, there is an extended plateau of values near $\log M_{\rm b}\sim 8$.
$\log M_{\rm b}\gtsim 8$ is found when very few of the 38 galaxies with $M_{\rm BH}> 10^7\msun$ are included in a bootstrap realisation of the data set.
In such cases, none of the data constrains $M_{\rm b}$ other than being larger than the largest $M_{\rm BH}$, and so $M_{\rm b}$ can take any arbitrarily large number without affecting how well the model fits the data.
This is also responsible for the very extended distribution of $k_2$ seen in Fig. \ref{fig:model1covarsfr}.

Function $f_3$ produces fairly consistent values with those of $f_1$, favouring a slightly higher value $\log M_{\rm b}\approx7.6$, as well as predicting a wide range of black hole masses over which AGN feedback quenches star formation with $\ldm\sim1 - 2$.
Similar caveats apply to $k_2$, $\log M_{\rm b}$, and $\ldm$ in this model as for $k_2$ and $\log M_{\rm b}$ in $f_1$.
Fortunately, however, the conclusions we draw in subsequent sections do not depend on the exact values of these parameters.

\begin{figure}
	\centering
	\includegraphics[width=0.48\textwidth,keepaspectratio]{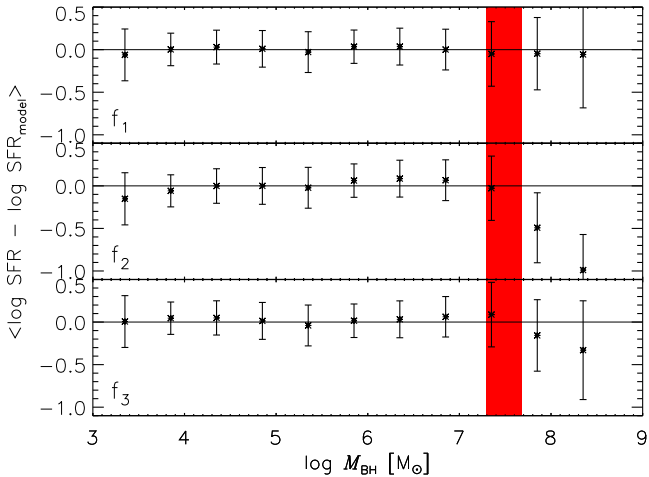}
	\caption{Residuals for the three models (assuming the best fit parameters) presented in Section \ref{sec:model}, as a function of BH mass.
	The mean and standard deviation of the residuals in bins of $\log M_{\rm BH}$ with a width of 0.5 dex are shown.
	From top to bottom, the panels show this quantity for $f_1$, $f_2$, and $f_3$, respectively.
	The red band indicates the range of $M_{\rm b}$ from Table \ref{tab:sfrfits}; $f_2$ clearly does not fit the data at $M_{\rm BH}>M_{\rm b}$.}
	\label{fig:resids}
\end{figure}
Fig. \ref{fig:resids} shows the residuals for each of the models $f_1$ to $f_3$ as a function of $M_{\rm BH}$.
The mean and standard deviation are shown for bins with width 0.5 dex, and the red band shows the range of $M_{\rm b}$ from Table \ref{tab:sfrfits}.
At $\log M_{\rm BH}\ltsim7$, the three models have residuals that are distributed fairly uniformly around 0, though $f_2$ may show a very slight increasing trend with $M_{\rm BH}$.
At higher masses, $f_2$, which assumes that there is no break mass $M_{\rm b}$, shows a significant systematic decrease in residuals with mass that is not seen in the residuals of $f_1$.
A similar trend is seen for the residuals of $f_3$, though the residuals are consistent with 0 over the full range of $M_{\rm BH}$.

\section{Discussion}\label{sec:discussion}
\subsection{BH growth and AGN feedback}

\begin{figure}
	\centering
	\includegraphics[width=0.48\textwidth,keepaspectratio]{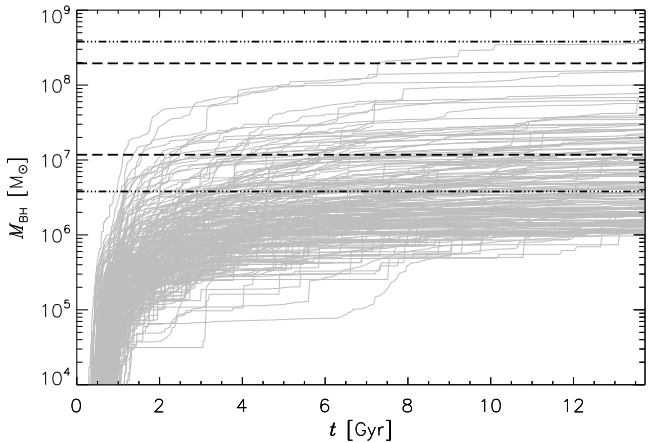}
	\caption{Mass assembly history of all BHs that grow above $10^6\msun$ by $z=0$, as a function of the age of the Universe.}
	\label{fig:ks_bh_mbhz}
\end{figure}
We focus now on understanding the implications of the value of $\ldm$ from $f_3$.
It is instructive to estimate a timescale over which BHs grow by a factor $10^{\ldm}\sim 10-100$.
In Fig. \ref{fig:ks_bh_mbhz} we show the mass assembly history of all simulated BHs with final masses greater than $10^6\msun$, and the horizontal dashed and dot-dashed lines show $M_{\rm BH}=M_{\rm b}\times10^{\pm\frac{1}{2}\ldm}$ assuming the best-fit and modal values, respectively.
It is clear that, regardless of the exact values of $M_{\rm b}$ and $\ldm$ used, the timescale for BHs to grow by a factor $10^{\ldm}$ is at least several Gyr.
This should not be interpreted, however, as the timescale over which an individual BH quenches star formation in its host, since we showed in \citet{pt15a} that star formation in individual galaxies can be quenched abruptly.
Rather, this shows that once they grow to $\sim M_{\rm b}$, BHs have the potential to exert influence over the evolution of their galaxy.

The fact that an individual BH can quench star formation in its host much more quickly than it takes to grow by a factor $10^{\ldm}$ once it reaches $M_{\rm BH}\approx M_{\rm b}$ implies that attaining a mass $M_{\rm b}$ is a necessary but not sufficient requirement for AGN feedback to become effective.
This may suggest that local properties within a galaxy can affect exactly when AGN feedback shuts off star formation, or that a `trigger' for strong AGN feedback may be required once the BH is sufficiently massive, with gas-rich galaxy mergers being a likely candidate \citep[e.g.,][]{comeford15,gatti16}.
The cause of strong AGN activity, such as the AGN-driven outflows described in \citet{pt15b}, will be investigated in detail in a future work.

\subsection{Galaxy transition at $3\times10^{10}\msun$}

Figure \ref{fig:resids} shows that the model that does not include a BH break mass, $f_2$, does not describe our simulated data well for $M_{\rm BH}\gtsim M_{\rm b}$.
Therefore we can conclude that the influence of BHs on star formation changes once the BH grows above $M_{\rm b}$.
There is an empirical relationship between $M_{\rm BH}$ and the stellar bulge mass of its host galaxy, $M_{\rm bulge}$.
Converting the range of fitted values  of $M_{\rm b}$ from both $f_1$ and $f_3$ ($M_{\rm b}\approx2-5\times10^7\msun$) into a stellar bulge mass using the relation given in \citet{grahamscott15} gives $M_{\rm bulge}\approx 1-2\times 10^{10}\msun$.
At these stellar masses, galaxies typically have bulge-to-disc ($B/D$) ratios of $0.3-0.4$ \citep{shen03}, corresponding to bulge-to-total ($B/T$) ratios of $0.23 - 0.29$ \citep[note that ETGs can have much larger $B/T\sim0.7$; see, e.g.,][]{morselli17}.
For such values of $B/T$, $M_{\rm b}$ corresponds to a galaxy mass that is consistent with the galaxy mass of $M_*=3\times 10^{10}\msun$ found by \citet{kauffmann03} to separate ETGs and LTGs.

This consistency between our simulations and observational data implies that feedback from super-massive BHs may be responsible for the transition from LTGs to ETGs at $M_*=3\times10^{10}\msun$.
Such a result is in broad agreement with other studies; \citet{keller16} find that even very strong supernova feedback in the form of superbubbles is inefficient in galaxies $M_*\gtsim4\times10^{10}\msun$, arguing that AGN feedback must dominate in more massive galaxies.
Similarly, \citet{bower17} find that BH accretion is suppressed by supernova feedback in galaxies less massive than $3\times10^{10}\msun$, with AGN feedback becoming important once supernova feedback can no longer remove gas from the galaxy potential.
In this paper, we have not investigated \emph{why} strong AGN feedback is triggered in massive galaxies, finding only that it \emph{is} triggered once a BH grows to $M_{\rm b}\approx2-5\times10^7\msun$.
The reason for the triggering will be investigated in detail in a future work.

 {
\subsection{Connection to the Kennicutt-Schmidt Law}
The Kennicutt-Schmidt (KS) law is an empirical relation between SFR surface density $\Sigma_{\rm SFR}$, and gas surface density, $\Sigma_{\rm gas}$ \citep{schmidt59,kennicutt89,kennicutt12}.
With $\gamma\approx1$ (the exponent of $M_{\rm gas}$ in our model), dividing equation \eqref{eq:10tomodel} by galaxy area gives $\Sigma_{\rm SFR}\sim\Sigma_{\rm gas}$.
Observations typically find a super-linear relation when all gas is considered \citep[e.g.,][]{kennicutt89,kennicutt98}, as we have done, while a linear or sub-linear relation tends to be found when only dense gas is considered \citep[e.g.,][but see also \citealt{liu11,momose13}]{bigiel08,shetty13,shetty14l}.
\citet{shetty14l} analysed disk galaxies from the STING survey and concluded that galactic properties other than $\Sigma_{\rm gas}$ affect $\Sigma_{\rm SFR}$; such properties were included in our model, which may explain the difference between our linear relationship and observations \citep[though there is significant scatter in the KS relation; see e.g.][]{krumholz12,cf13,salim15}.
We note for completeness that if we only consider SFR and $M_{\rm gas}$, we obtain a slope $1.54\pm0.03$, in much better agreement with the observed super-linear relation of \citet{kennicutt89,kennicutt98}.
}

 {
\subsection{The star formation main sequence}
Regardless of the form of $f(M_{\rm BH})$ used, Table \ref{tab:sfrfits} shows that $\alpha=0.21$, i.e. $\log {\rm SFR}\sim 0.21\log M_*$.
This is in contrast to observational estimates, which find a present-day slope around 0.75 -- 1 \citep[e.g.,][]{elbaz07,elbaz11,zahid12,renzini15}.
This is due to the correlations between stellar mass and other physical properties that are not accounted for in the standard SFMS, but are in this model, as was the case for the KS relation above.
Our results suggest that stellar mass is not as important in determining galactic SFR as implied by the standard SFMS, and that there is an environmental dependence as well as strong gas mass dependence on the SFR of star-forming galaxies.
}

 {
\section{Comparison of our new SFR Model to observational data}\label{sec:obs}
We test our star formation models by comparing to observations.
Ideally, we would wish to repeat the analysis of Section \ref{sec:model_disc} on observational data to critically appraise our star formation model and our simulations.
However, there are no overlapping surveys that can provide all of SFR, $M_*$, $s_5$, $M_{\rm gas}$, and $M_{\rm BH}$ for a large number of galaxies.
By searching the literature and various current observational databases, we were able to find a sample of galaxies for which all the required measurements of SFR, $M_*$, $s_5$, and $M_{\rm gas}$ were available (except $M_{\rm BH}$).
We use this set of galaxies to compare their measured SFR with that estimated from our model.
}

 {
The Galaxy And Mass Assembly survey \citep[GAMA,][]{GAMA} is an optical survey using the Anglo-Australian Telescope and AAOmega spectrograph \citep{sharp06}.
It is $>$98 per cent complete to a depth of $r<19.4$ or $19.8$.
The survey is spread over three $12^\circ$ by $4^\circ$ equatorial regions.
Environmental density data are only available for 16,062 galaxies in the region centred at 15h right ascension.
In this paper, we use aperture-corrected $M_*$ \citep{taylor11}, SFR \citep{gunawardhana11}, and $s_5$ \citep{brough13} data for these galaxies.
}

 {
For the gas masses, we use data from the Arecibo Legacy Fast ALFA survey \citep[ALFALFA,][]{ALFALFA}.
ALFALFA is a wide-field survey using the Arecibo radio telescope to map the 21 cm line of atomic hydrogen.
Corresponding HI masses, $M_{\rm HI}$ are derived in \citet{haynes11}; to convert this to total gas mass, we make the simplistic assumptions that 1) $M_{\rm HI}$ is equal to the total mass of hydrogen in the ISM, and 2) the ISM has primordial composition, so that $M_{\rm gas}=M_{\rm HI}/0.75$.
We associate HI detections from ALFALFA with an optical counterpart in GAMA if they are separated by less than $0\fdg01$ in RA and Dec.
Matching the catalogues in this way gives 14 galaxies with measured $M_*$, SFR, $s_5$, and $M_{\rm gas}$.
}

 {
\begin{figure}
	\centering
	\includegraphics[width=0.48\textwidth,keepaspectratio]{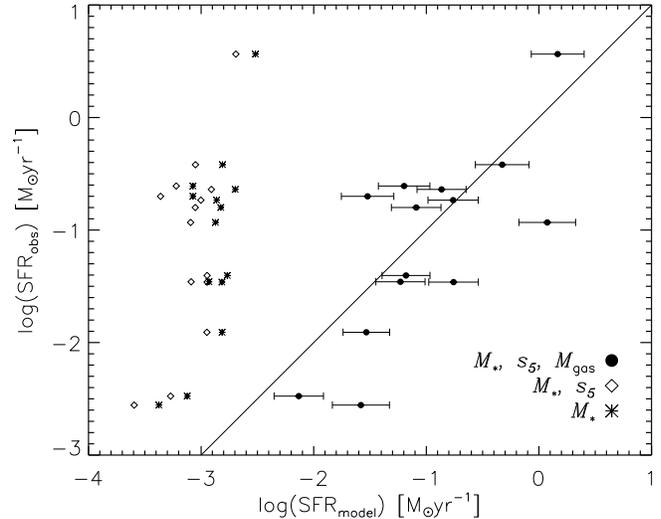}
	\caption{ {Comparison of measured SFR (SFR$_{\rm obs}$) and predicted SFR (SFR$_{\rm model}$) from our model, using the best-fit parameters of Model 1 (equations \eqref{eq:model} and \eqref{eq:model1}, and Table \ref{tab:sfrfits}), for galaxies selected from the GAMA and ALFALFA surveys.
	The solid line denotes SFR$_{\rm model} = {\rm SFR}_{\rm obs}$.
	Filled circles with error bars show the results of equation \eqref{eq:obsmodel}, while diamond and star symbols are for if $M_{\rm gas}$ or $M_{\rm gas}$ and $s_5$ are excluded, respectively.
	Error bars on SFR$_{\rm model}$ include both errors on observed quantities and uncertainties in the model parameter values; no error information was available for SFR$_{\rm obs}$.}}
	\label{fig:sfrobs}
\end{figure}
We estimate the SFR of these galaxies assuming the best-fit parameters of Model 1, neglecting any contribution from BHs, i.e.,
\begin{equation}\label{eq:obsmodel}
\begin{split}
	\log & \left({\rm SFR}_{\rm model}\right) = -2.50 + 0.21\log\left(\frac{M_*}{10^{10}\msun}\right) \\
	& - 0.17\log\left(\frac{s_5}{100\,{\rm kpc}}\right) + 1.03\log\left(\frac{M_{\rm HI}}{0.75\times10^7\msun}\right).
\end{split}
\end{equation}
The most massive of these 14 galaxies has $M_*=1.7\times10^{10}\msun$, and so we expect to be in the parameter space where BH feedback does not play a crucial role for the SFR, i.e. $\log f\sim k_1M_{\rm BH}\sim0$.
Higher-mass galaxies with $M_*$, $s_5$, $M_{\rm gas}$, and simultaneous $M_{\rm BH}$ measurements are needed in the future to test our predicted dependence of the SFR on BH mass.
}

 {
In Fig. \ref{fig:sfrobs} we show the measured SFR (SFR$_{\rm obs}$) and the SFR predicted by our model (SFR$_{\rm model}$) for these galaxies (filled circles with error bars).
The agreement between the observed SFR and those predicted by our model is extremely encouraging, considering the assumptions used to calculate $M_{\rm gas}$ and the fact that $s_5$ from the GAMA catalogue is projected, whereas we used the 3-D $s_5$ in deriving our model parameters.
Also shown in Fig. \ref{fig:sfrobs} are the predicted SFRs if $M_{\rm gas}$ or $M_{\rm gas}$ and $s_5$ are excluded (diamond and star symbols, respectively).
In both cases, the model SFR is significantly less than is observed, and shows little evidence of correlation with the observed values.
This reinforces the fact that $M_{\rm gas}$ is the most important quantity in determining SFR.
The inclusion of $s_5$ alters the predicted SFR very little, but may be more important in more extreme environments such as galaxy groups and clusters \citep[e.g.,][]{schaefer17}.
}

 {
In the coming decade, vast radio surveys such as the Square Kilometre Array (SKA) and its precursors will provide HI masses for tens of thousands of galaxies, allowing for a more rigorous test of our model.
However, black hole masses are more difficult to measure, and are available for only a relatively small number of galaxies in the local Universe.
Directly validating our full model including the effects of BHs will be a challenging, but important task for the near future.
}


\section{Conclusions}
\label{sec:conc}

We have analysed data from a cosmological hydrodynamical simulation that includes a detailed prescription for star formation and stellar feedback, as well as BH formation and AGN feedback.
Our aim was to understand better the relative importance of factors that affect SFR of galaxies.
We suggested that SFR could be influenced by the mass of the host galaxy, its environmental density, its gas content, and the mass of its BH.
We proposed a relatively simple form for the relation between SFR and host galaxy properties (equation \eqref{eq:model}), with which we were able to make quantitative comparisons.

We find that once a BH has grown sufficiently massive ($\sim2-5\times10^7\msun$), it can be triggered to shut off star formation through AGN feedback; the details of such triggering will be investigated in future work.
Such a BH mass corresponds, via the Maggorian relation \citep{grahamscott15}, to a galaxy bulge mass of $\sim 1-2\times10^{10}\msun$, which is consistent with the observed galaxy mass above which ETGs make up the dominant fraction of the galaxy population \citep[$\sim 3\times10^{10}\msun$,][]{kauffmann03}.
It is important to note that these masses are emergent in our simulations; none of the input parameters of the baryon physics is chosen to yield such a transition mass, but are constrained by other observations.

The SFR depends strongly on the amount of gas in a galaxy, scaling approximately linearly with gas mass,  {while the dependence on stellar mass is weaker than predicted from the standard SFMS since we take into account the correlations between SFR and the additional physical quantities $M_{\rm gas}$, $s_5$, and $M_{\rm BH}$ by fitting all simultaneously.}
Our simulation cannot resolve the cold molecular gas that would form stars, and in this analysis we have treated all gas equally, regardless of temperature or density.
In light of this, it will be useful in future works to analyse galaxies simulated with sufficient resolution to distinguish between gas phases.

We find that star-forming galaxies in high density regions have larger SFR than star-forming galaxies in the field at given mass, in agreement with observations \citep[e.g.,][]{koyama13}.
This would most likely evolve with redshift, with massive galaxies showing the highest SFR in the past.
The existence and strength of any such evolution will be investigated in a future work.
Furthermore, the size of our simulation box limits the size of our most massive simulated cluster, and it would be informative to investigate if the trends seen here hold for more extreme environments.

 {
We use observational data from the GAMA and ALFALFA surveys to compare the SFR observed with that predicted by our new model (see Fig. \ref{fig:sfrobs}).
There is excellent agreement between these two quantities, lending credence to the model we have presented.
The sample is limited by the available observations, but will increase to tens of thousands in the era of the SKA.
Further validation of our model will only be possible with a catalogue of BH masses for large numbers of galaxies.
}


\section*{Acknowledgements}

 {We thank the referee for their useful comments, which improved the quality of this paper.}
C.F.~gratefully acknowledges funding provided by the Australian Research Council's Discovery Projects (grants~DP150104329 and~DP170100603).
The simulations presented in this work used high performance computing resources provided by the Leibniz Rechenzentrum and the Gauss Centre for Supercomputing (grants~pr32lo, pr48pi and GCS Large-scale project~10391), the Partnership for Advanced Computing in Europe (PRACE grant pr89mu), the Australian National Computational Infrastructure (grant~ek9), and the Pawsey Supercomputing Centre with funding from the Australian Government and the Government of Western Australia, in the framework of the National Computational Merit Allocation Scheme and the ANU Allocation Scheme.
This work has made use of the University of Hertfordshire Science and Technology Research Institute high-performance computing facility.
PT thanks S.~Lindsay for helpful discussions.
Finally, we thank V.~Springel for providing {\sc GADGET-3}.


\bibliographystyle{mn2e}
\bibliography{./refs}



\bsp

\label{lastpage}

\end{document}